\documentclass{PoS}
\usepackage{graphicx}
\usepackage{here}

\def\beq{\begin{equation}}
\def\eeq{\end{equation}}
\def\bea{\begin{eqnarray}}
\def\eea{\end{eqnarray}}
\def\ba{\begin{array}}
\def\ea{\end{array}}
\def\bq{\begin{quote}}
\def\eq{\end{quote}}

\def\la{\langle}
\def\ra{\rangle}
\def\nin{\noindent}

\def\als{\alpha_s}

\def\g2{ \la\alpha_s G^2 \ra}
\def\g3{g^3f_{abc}\la G^aG^bG^c \ra}
\def\g4{\la\als^2G^4\ra}

\title{Tribute to Francisco (Paco) Yndurain}
\ShortTitle{Paco Yndurain}

\author{
\speaker
{Stephan Narison}
\thanks{Invited plenary talk.}
 \\
        Laboratoire de Physique Th\'eorique et Astroparticules, CNRS - IN2P3 \& Universit\'e
de Montpellier II, Case 070, Place Eug\`ene Bataillon, 34095 - Montpellier Cedex 05, France\\
        E-mail: \email{snarison@yahoo.fr}}

\abstract{This part of the talk aims to present briefly the biodata and the exceptional career of
Paco Yndurain  who left us suddenly  in June 2008. 
The scientific part: {\it Light Scalar Mesons in QCD} is published in the proceedings of QCD 08 (Montpellier 7-12th july 2008: arXiv:0811.0563 [hep-ph]). }

\FullConference{8th Conference Quark Confinement and the Hadron Spectrum \\
		 September 1-6 2008\\
		 Mainz, Germany}

\begin{document}

\section{Introduction}
\nin
These last months were very sad for our community as we have suddenly lost our colleagues and friends:
{\begin{figure}[here]
{\includegraphics[width=4.5cm]{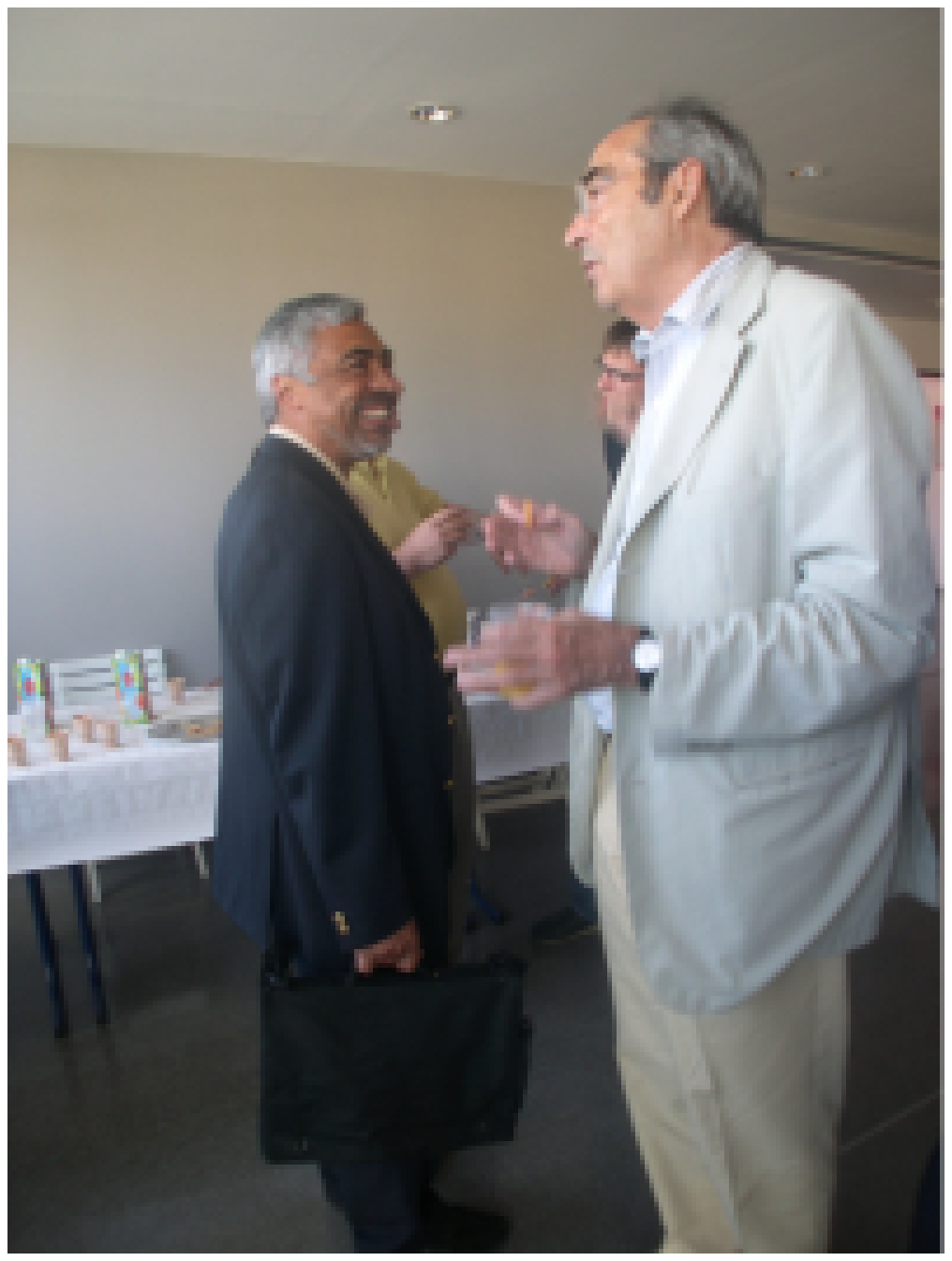}}
{\includegraphics[width=5.5cm]{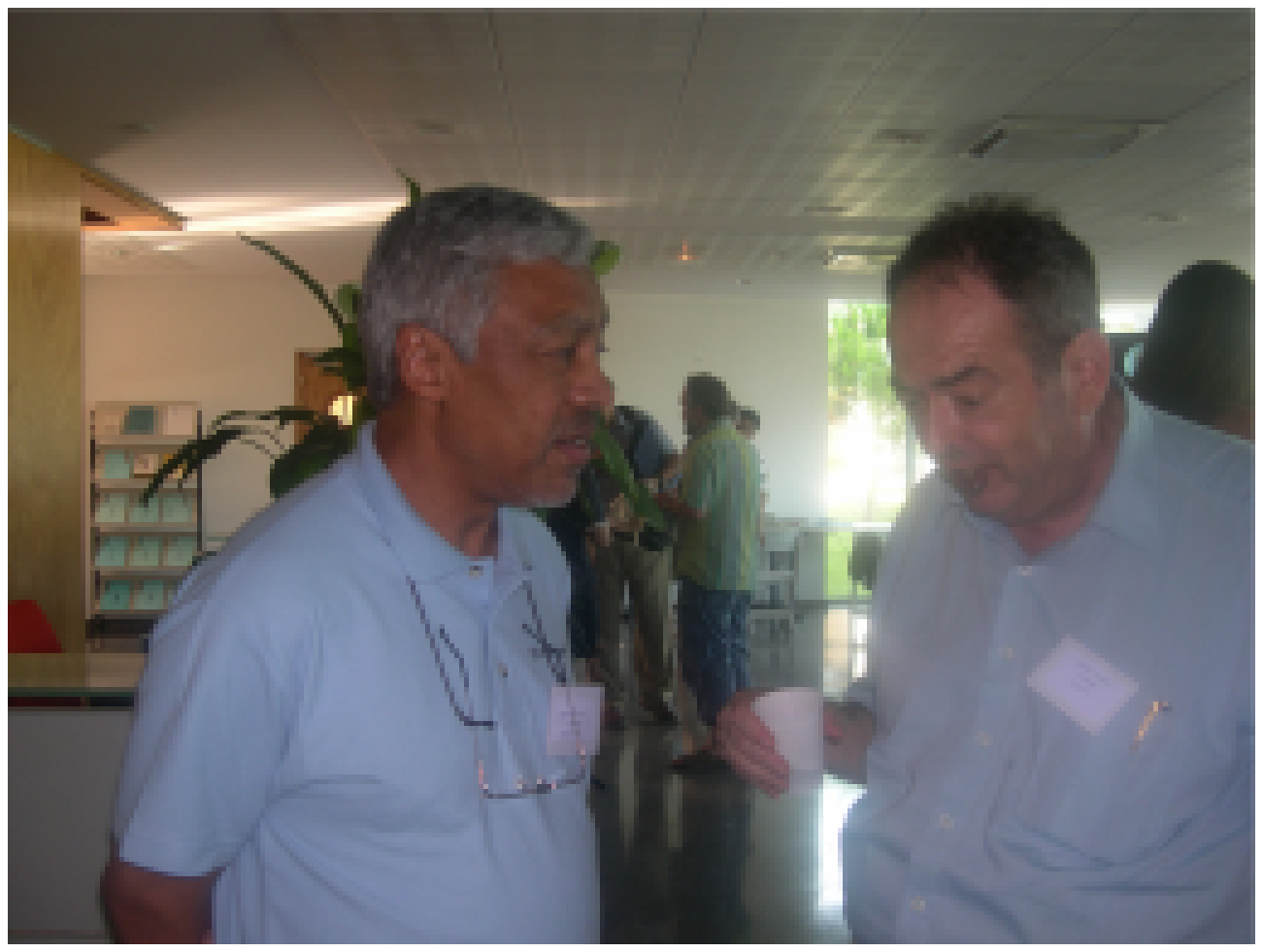}}
\includegraphics[width=2.5cm]{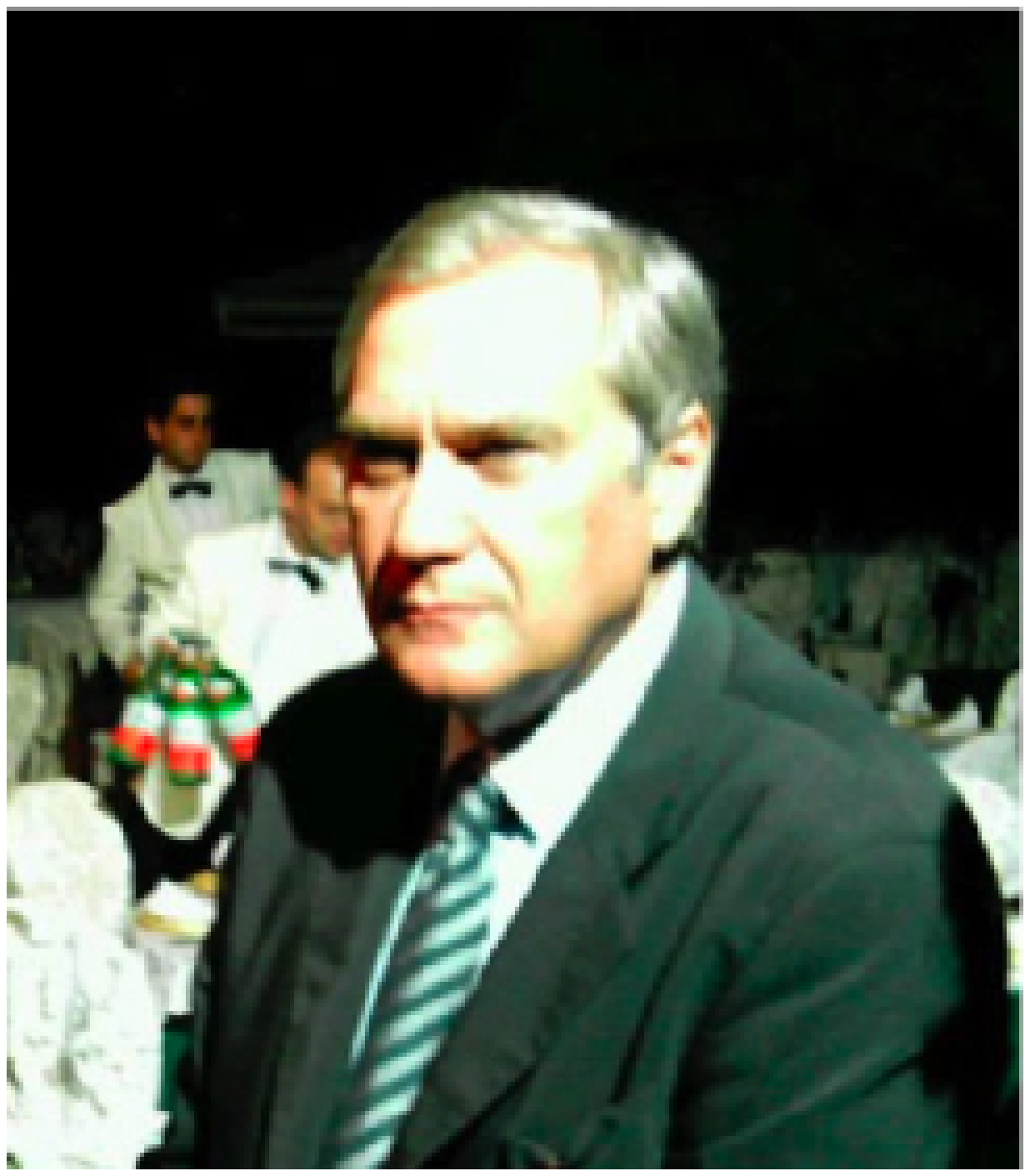}
\end{figure}} 

\nin
{\it 
Paco YNDURAIN (Dec. 1940 - June 2008), Jan STERN (June 1942 - July 2008) and \\
Giuseppe NARDULLI (July 1948 - June 2008) }\footnote{The two first photos have been taken during Paco
and Jan participation at QCD 06 (Montpellier, 4-7th July 2006).}.\\

\nin
{\it Before starting my talk dedicated to Paco Yndurain, I ask 1mn of silence for their memory.}

\section{Introduction to Paco}
\begin{figure}[here]
\hspace*{4cm}{\includegraphics[width=6.80cm]{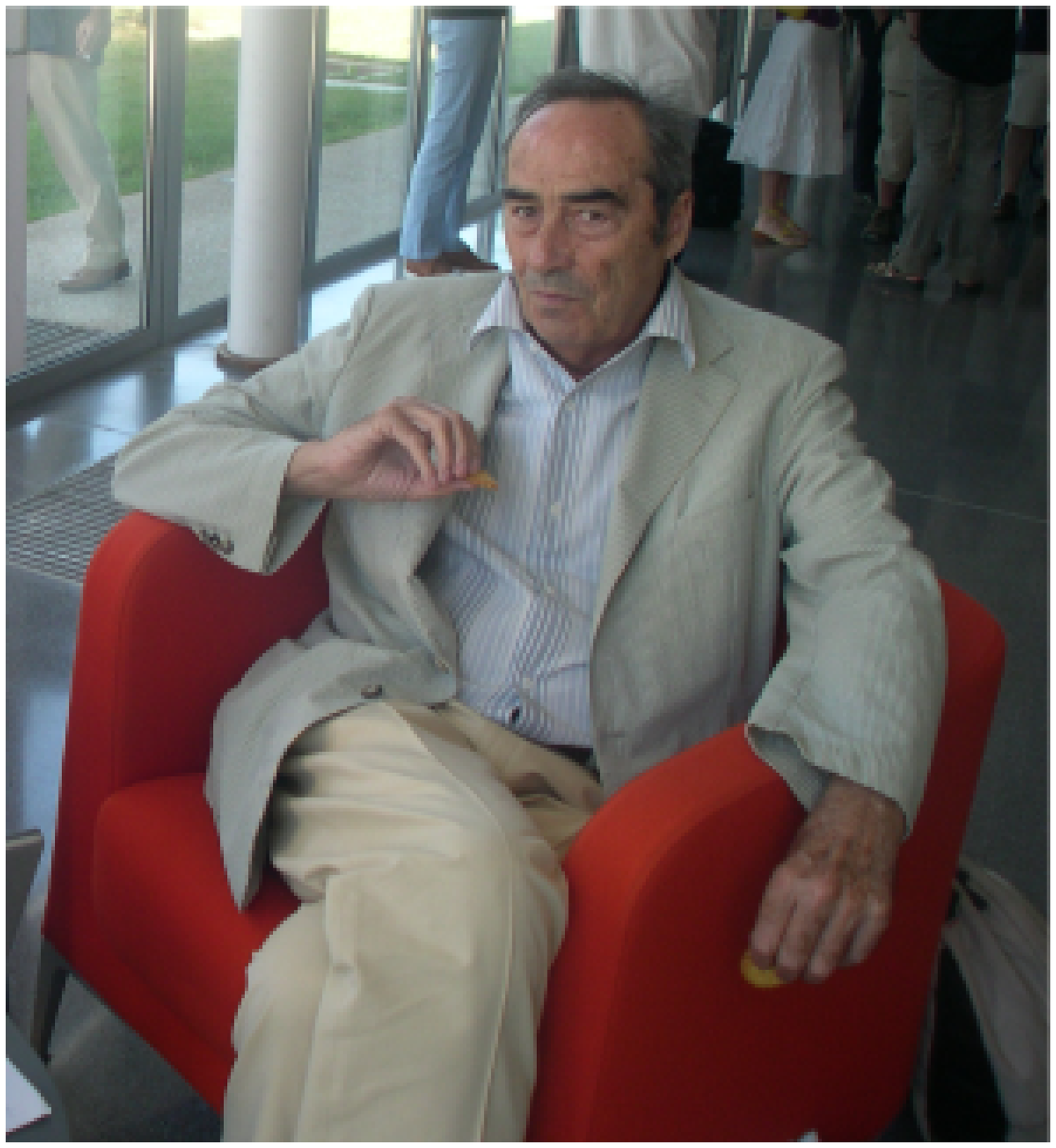}}
\end{figure}
\hspace*{5.5cm}{\it Paco at QCD 06}
\\
\\
\nin
I met Paco in 1975 when preparing my ``Doctorat of 3\`eme cycle" in Marseille on the ``hadronic contributions to the muon anomalous magnetic moment'', where he has been one of the referees of my thesis. In that time,  I have been impressed by his physics intuition and knowledge and by his friendship. I have appreciated his human relation and also benefited from his  encouragement during his multiple visits to Marseille. Later on in 1979, we have started to work on the ``sum of light the quark masses in QCD" with Carlo Becchi and Eduardo de Rafael.  He has invited me to the Pyrenees school and then to Madrid in 79. Driving to Madrid, he has friendly convinced me to compute the radiative corrections to the (pseudo)scalar two-point correlator which few months later becomes the first calculation of this quantity and the first prediction on the running values of the sum of light quark masses in QCD. Since then, though we have not worked directly together, we have continued to share our interests in physics (especially in QCD). Paco was also the International Committee member of the QCD Conference series in Montpellier since its creation in 1985, where he came often as an active participant. \\
In the following, I shall give a short review of his formation and brilliant career \footnote{Some parts of this talk have been inspired from the one of Jos\'e Pelaez at QCD 08 ( 7-12th July 2008, Montpellier).}.
\section{Paco formation and itinerary }
\nin
{\it \bf University of Zaragoza} 
 \begin{itemize}
{\small  \item  Master in Mathematics 1962, 
 \item PhD in Physics 1964: ``Definitions of hamiltonians and renormalization"
\item  Assistant Prof. 64-66}
    \vspace*{-3.5cm}
    {\begin{figure}[here] \flushright
{\includegraphics[width=3.5cm]{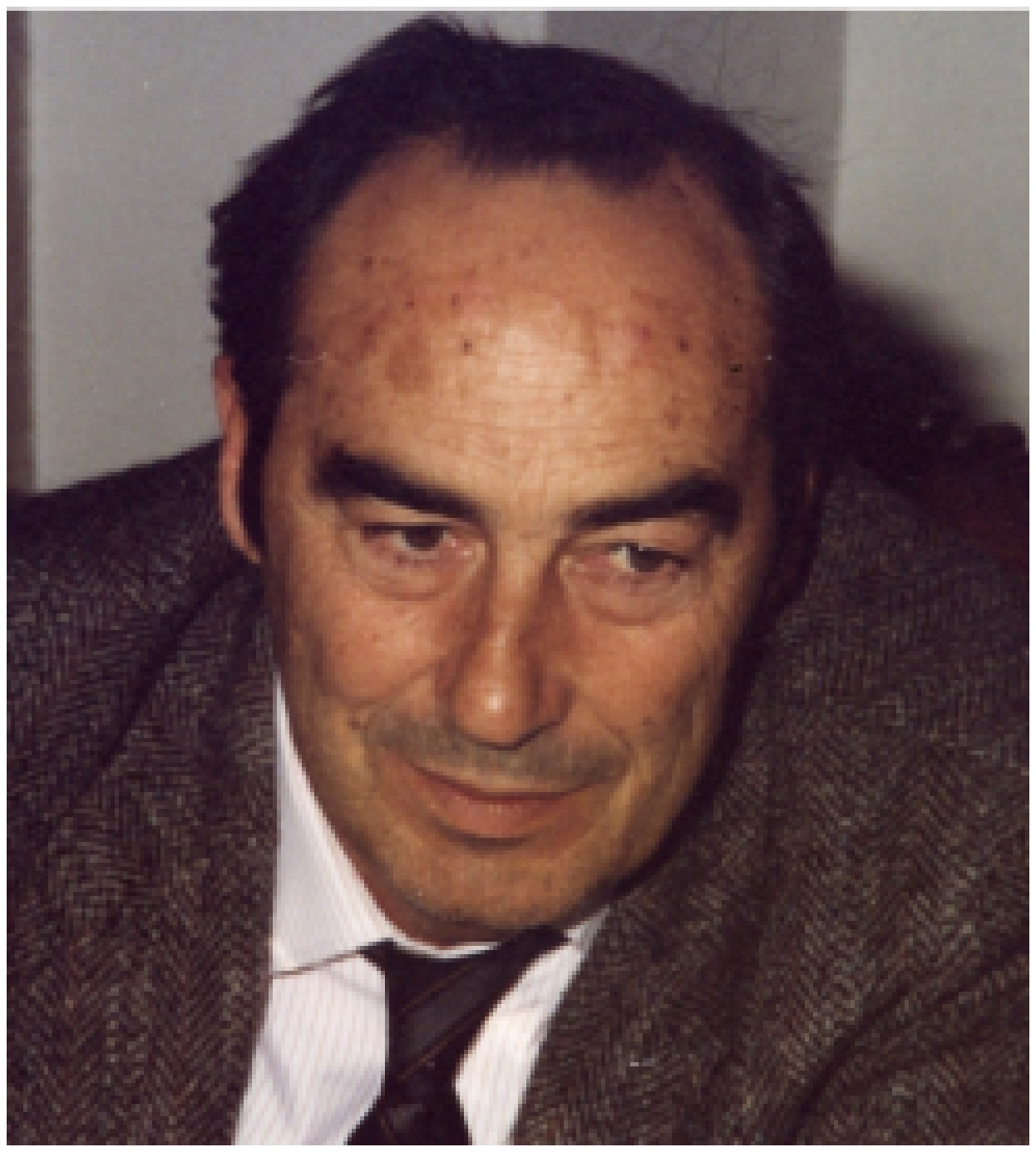}}
\end{figure}} 
\vspace*{-1cm}
\end{itemize}
{\it \bf University of New York }
 \begin{itemize}
{\small  \item Fullbright Fellow 66-67
\item  Associate Researcher 67-68}
\end{itemize} 
{\it \bf CERN Theory Division}
 \begin{itemize}
{\small  \item  Research Fellow 68-70
\item Senior Scientific Associate 76 and 85}
 \end{itemize} 
 {\it \bf University of Madrid}
 \begin{itemize}
{\small  \item 
 Professor 1970 U. Complutense
\item  Full Professor since 1971 in the Aut\`onoma University.}
   \end{itemize} 
\centerline{ \it  Paco is reputed to be an Excellent Teacher !}
\section{PACO  (important) positions }
 \begin{itemize}
{\small \item  Director of the Theoretical Physics Department UAM, 74-77, 81-84
 \item Dean of the UAM Science Faculty 1975
 \item Research Vice-rector UAM 78-81  
 \item Scientific Advisor of IBM 83-85 and of Scientific Research of Kuwait 80-82
 \item Elected Member of the Eur. Phys. Soc., 83-89
 \item Elected Member of the Scientific Policy Committee of CERN, 88-94.}
  \vspace*{-6.cm}
{\begin{figure}[here] \flushright
{\includegraphics[width=3cm]{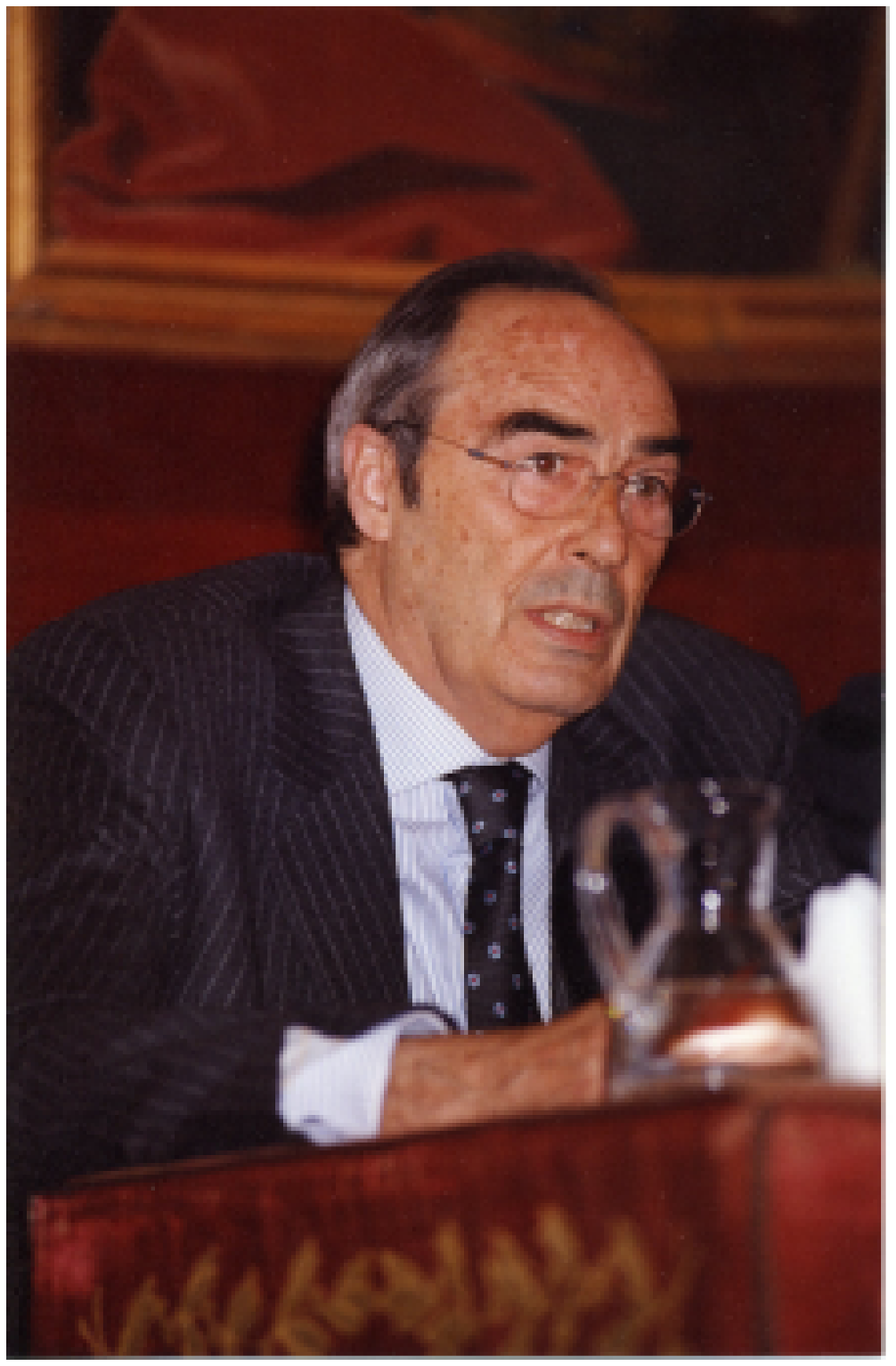}}
\end{figure}} 
  \vspace*{0.5cm}
{\small \item Elected Member by the Senate of the Spanish University Council
 \item Elected Member of the World Innovation Foundation, London, 2001
 \item Elected President of the Physics and Chemistry Section of the Spanish Royal Academy 2002.}

 \end{itemize} 
\centerline{ \it Among the Fathers \footnote{It includes my former collaborator and friend Pedro Pascual who also left us few years ago.} of  High-Energy Physics (HEP)  in Spain !}
\section{PACO  Mobilities }
 \begin{itemize}
{\small \item  CERN - Theory division
 \item  Michigan { (Veltman-Zakharov)}
 \item Marseille {(de Rafael)}, 
 \item NIKHEF, Brookhaven, Orsay, Saclay, Groningen, Vienna, La Plata, Bogot\`a, Caracas.}
\end{itemize}
\section{PACO  Societies \& Honours}
 \begin{itemize}
{\small \item Founder member of the European Physical Society
 \item Invited member of the American Association for the Advancement of Science 
 \item Invited memeber of the New York Academy of Sciences
 \item Member of the Spanish Royal Physics Society
\item Member of the Spanish Royal Mathematical Society.}
\end{itemize} 
\section{International  Committee Member and Participation to Conferences}
 \begin{itemize}
{\small \item Series of Confinement : since  its creation in 1994
  \item Series of QCD-Montpellier: since its creation in 1985.}
  \end{itemize} 
  \section{PACO  (wide)  Research Interests: an Outstanding Physicist ! }
  \nin
 {\bf  Publications} \\
 Paco has worked in different fields of High-Energy Physics \\ 
 (electroweak, supersymmetry, QCD,
 pion-pion scattering,..). He has written:

  \begin{itemize}
{\small \item 158 publications (articles and conference talks)
 \item  Well-known QCD and QM books.}  
       \begin{figure}[here]
      {\includegraphics[width=3.2cm]{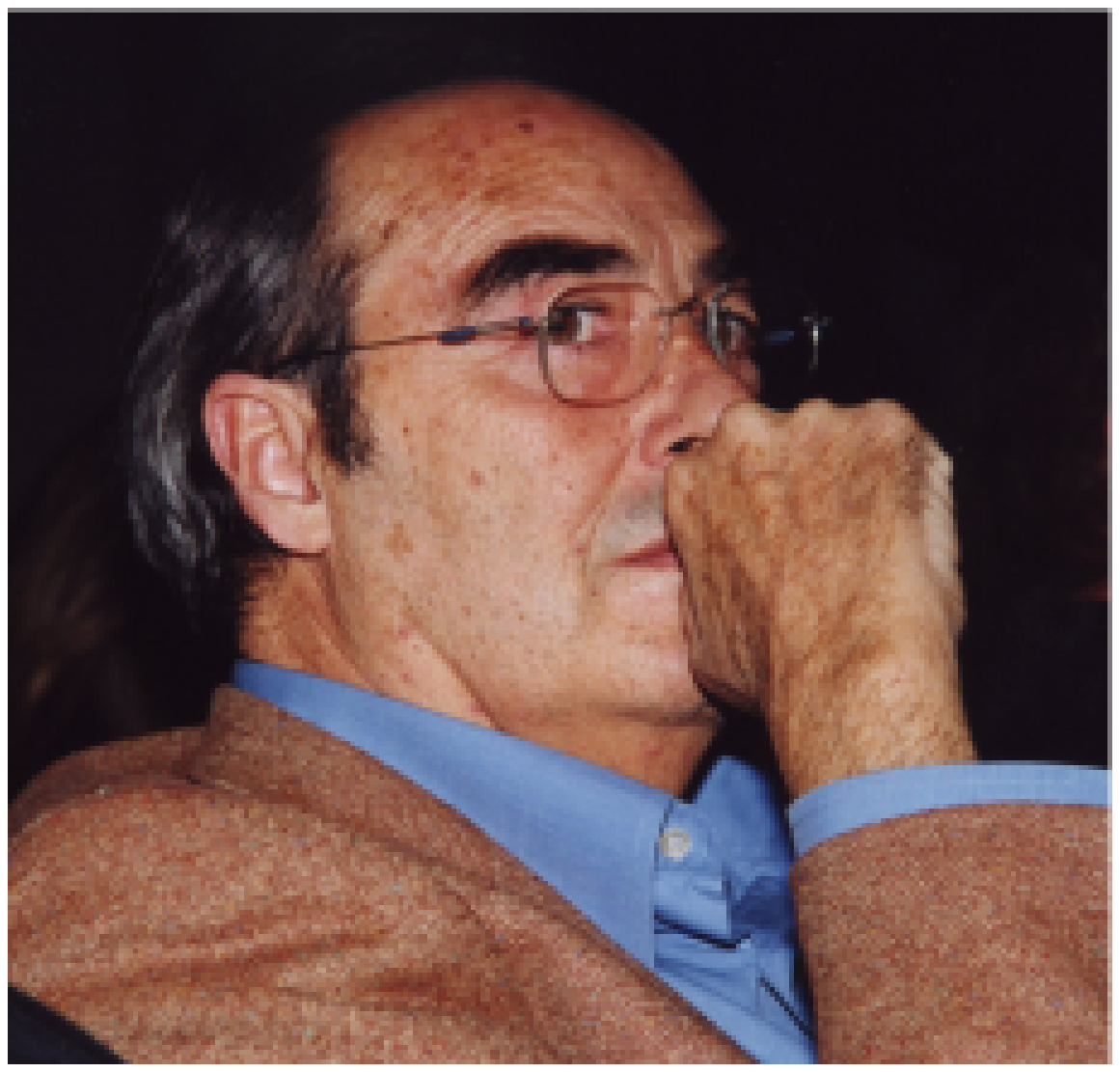}}
\hspace*{2.5cm}
{\includegraphics[width=3.2cm]{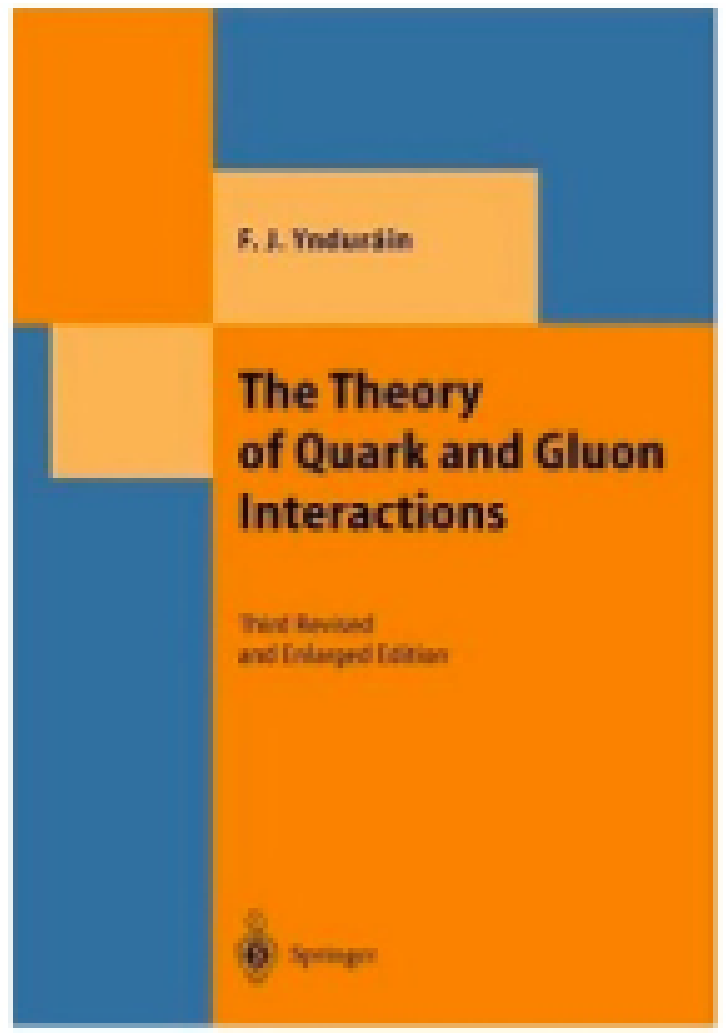}}
\hspace*{2.5cm}
{\includegraphics[width=3.05cm]{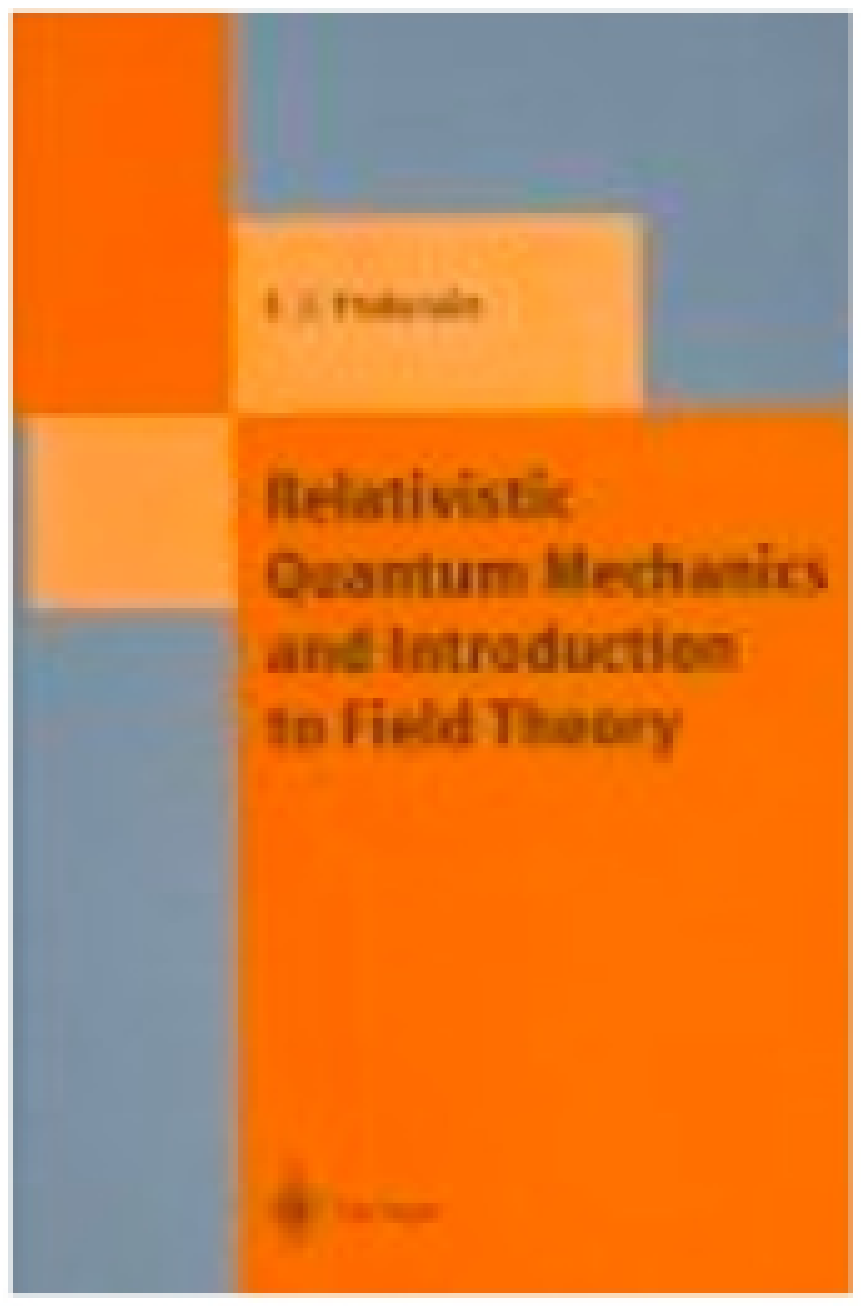}}
\end{figure}


\end{itemize}
I classify  below his different contributions to each fields in High-Energy Physics. 
The exact references of his papers can be e.g. found on-line in SLAC QSPIRES:\\
{\it http://www.slac.stanford.edu/spires/find/hep/wwwcite?rawcmd=FIND+AUTHOR+yndurain}
\\
\\
{\bf  EW Publications }
 \begin{itemize}
{\small \item 1979-1981 with C. Jarlskog \& De Labastida: Proton decay
 \item 1984 with Alvarez, Herrero, Ibanez \& Lopez :  SUSY
  \item 1983-85 with Cecilia Jarlskog :  \# of families \& space-time
    \item 1989 with Tini Veltman : rad. corr. in WW scattering.}
    \vspace*{-2.5cm}
    {\begin{figure}[here] \flushright
{\includegraphics[width=5cm]{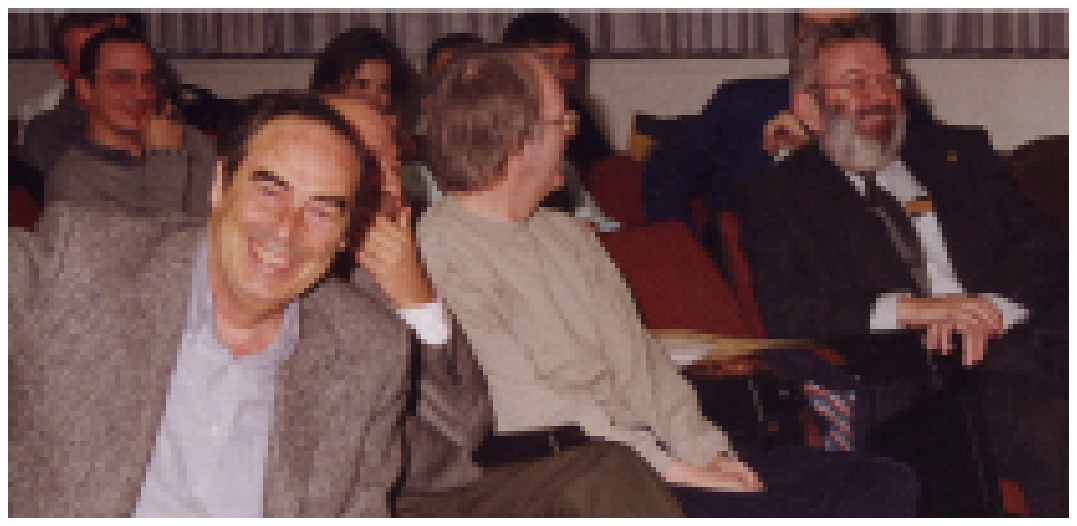}}
\end{figure}} 
\end{itemize}
 \vspace*{-.5cm}
{\bf QCD publications} \\
{\it Vector Meson Dominance}
 \begin{itemize}
{\small \item 1978-83 with Bernabeu, Bramon, Espriu, Tarrach.} 
\end{itemize} 
{\it Running Light Quark Masses}
 \begin{itemize}
{\small \item 1979-80 with Becchi, Narison \& de Rafael
 \item Other papers since 1980 and in 1982 with Gonzalez-Arroyo, Martinelli.}  
\end{itemize} 
{\it Deep Inelastic Scatterings }
 \begin{itemize}
{\small \item 1979-99 with Escoubes, Gonzalez-Arroyo, Herrero, Lopez \& Santiago : \\ NLO \& NNLO structure functions, $\alpha_s$ determination  
  \item 1997 with Adel, Barreiro, Labarga \& Vermaseren : Pomeron and small-x.}
\end{itemize} 
{\it Heavy Quarkonium properties}
 \begin{itemize}
{\small \item 1995-2000 with Pineda, Titard \& Simonov : $m_c$, $m_b$ and $\eta_b$}
\end{itemize} 
{\it Gluon condensate determination}
 \begin{itemize}
{\small  \item 1999 :  QCD spectral sum rules}
\end{itemize} 
{\it Hadronic Contributions to g-2 and $\alpha(M_Z)$}
 \begin{itemize}
{\small  \item 1983 with Casas \& Lopez
and 2001, 2005 with Troconiz
 \item Hot debate with Davier et al. }
     \vspace*{-5.5cm}
   {\begin{figure}[here] \flushright
{\includegraphics[width=3.5cm]{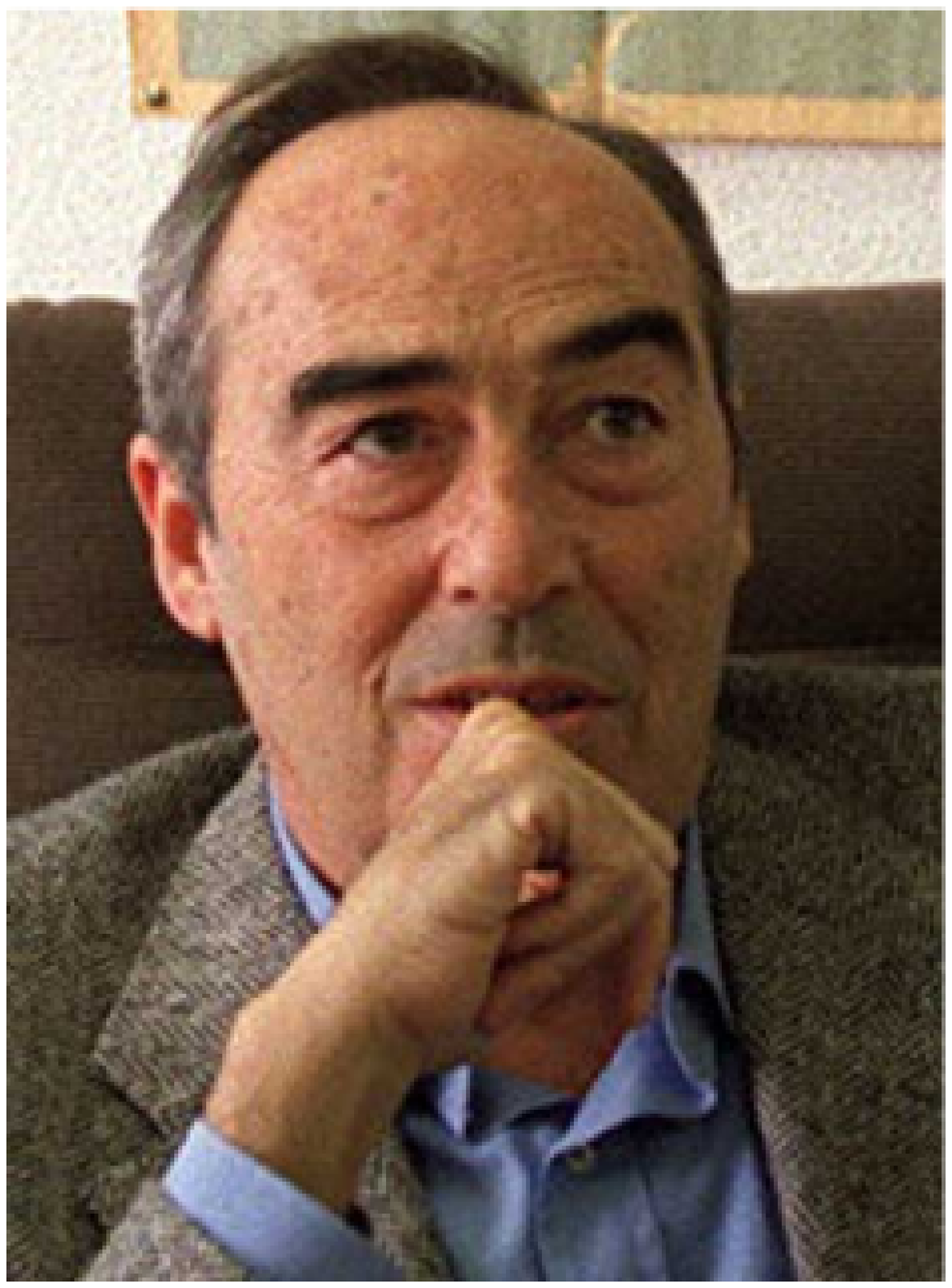}}
\end{figure}} 
\end{itemize} 
{\it Analyticity and $\pi\pi$ scatterings}
 \begin{itemize}
{\small \item Since 1969 with $\neq$ collaborators : Adler, Aguilar-Benitez, Atkinson, Cerrada, Common, Garcia-Martin, Gervais, Izquierdo, Johnson, Kaminski, Lopez,  Mahoux, Palou, P \& R. Pascual,
 Pelaez, Sanchez Gomez.
 \item Recent Hot Debate with Caprini, Gasser \& Leutwyler: 
 Paco has wished
 to close this debate in Montpellier but the Bern group did not come in 2006. The meeting has been planned for  this year 2008.
 { Paco has cancelled his participation to QCD 08 15 days before he left us ! }}
  \end{itemize} 
{\it  Comments and Famous Papers}\\
  Most of his papers  had an impact in the developments of the related fields.  In addition to his famous QCD book and to his recent work on $\pi\pi$ scattering reviewed in:
    \begin{itemize}
{\small  \item  In memory of Paco Yndurain: A Precise determination of pi pi scattering..., \\
J.R. Pelaez, R. Garcia-Martin, R. Kaminski, F.J. Yndurain, \\
QCD 08 (Montpellier  7-12th July 2008: arXiv:0810.2204 [hep-ph]),}
  \end{itemize} 
  his most quoted works in electroweak interactions are:
      \begin{itemize}
{\small \item
Matter Instability in the Su(5) Unified Model of Strong, Weak and Electromagnetic Interactions,\\
C. Jarlskog, F.J. Yndurain, Nucl. Phys. B149:29,1979.
\item Radiative Corrections to W W Scattering,\\
M.J.G. Veltman, F.J. Yndurain, Nucl. Phys. B325:1,1989. 
\item  How To Look For Squark with the Anti-P P Collider,\\
M.J. Herrero, Luis E. Ibanez, C. Lopez, F.J. Yndurain, Phys. Lett. B132:199-201,1983. }
  \end{itemize} 
 while his famous works in QCD are:
  \begin{itemize}
{\small \item The Hadronic contributions to the anomalous magnetic moment of the muon,\\
J.F. de Troconiz, F.J. Yndurain, Phys. Rev. D71:073008,2005 [hep-ph/0402285].
\item Calculation of quarkonium spectrum and m(b), m(c) to order $\alpha_s^4$,\\
A. Pineda, F.J. Yndurain, Phys.Rev.D58:094022,1998 [hep-ph/9711287].
\item Rigorous QCD evaluation of spectrum and ground state properties of heavy q anti-q systems: With a precision determination of m(b) M(eta(b)),\\
S. Titard, F.J. Yndurain, Phys.Rev.D49:6007-6025,1994 [hep-ph/9310236].
\item Light Quark Masses in Quantum Chromodynamics and Chiral Symmetry Breaking,\\
C. Becchi, Stephan Narison, E. de Rafael, F.J. Yndurain, Z.Phys.C8:335,1981.
 \item  Second Order Contributions to the Structure Functions in Deep Inelastic Scattering,\\
Antonio Gonzalez-Arroyo, C. Lopez, F.J. Yndurain, Nucl. Phys. B153:161-186,1979.  
\item Reconstruction of the Deep Inelastic Structure Functions from their Moments,\\
F.J. Yndurain, Phys.Lett.B74:68,1978. }
\end{itemize}
\section{Epilogue}
Why PACO not Fransisco ?
 \begin{itemize}
{\small \item { Saint Francisco} was the founding father of a monastic community.
 \item { PACO} was actually called the ``Father of the Community". ~ 
 In latin: ``{ PA}ter { CO} munalis" : {  Alvaro de Rujula} wrote { PAKO} which is stronger in Spanish !{ (from Eduardo de Rafael)}.
 \item Indeed, {Pako} was { among the Fathers} of the HEP Spanish Research
  and { the Father} of the Physics Department in the Autonoma Univ. of Madrid.
\item { Pako} also had { a huge influence} in the Spanish and International Physics Communities.
\item Pako was a ``typical Spanish man" for being professionally arrogant but a gentleman in every days life. }
\end{itemize} 
\section{Tribute to PAKO !}
 \begin{itemize}
{\small \item { \it His death is certainly a big lost to Physics  \\
 and especially to his Family 
 to whom \\ we address our sincere condolences !}
 \item \it However, Death is  like our Life, \\ a Natural Phenomena ! 
 \item { PAIX A TON AME :  PAKO !}}
\end{itemize} 
\vspace*{-4.5cm}
{\begin{figure}[here] \flushright
{\includegraphics[width=7cm]{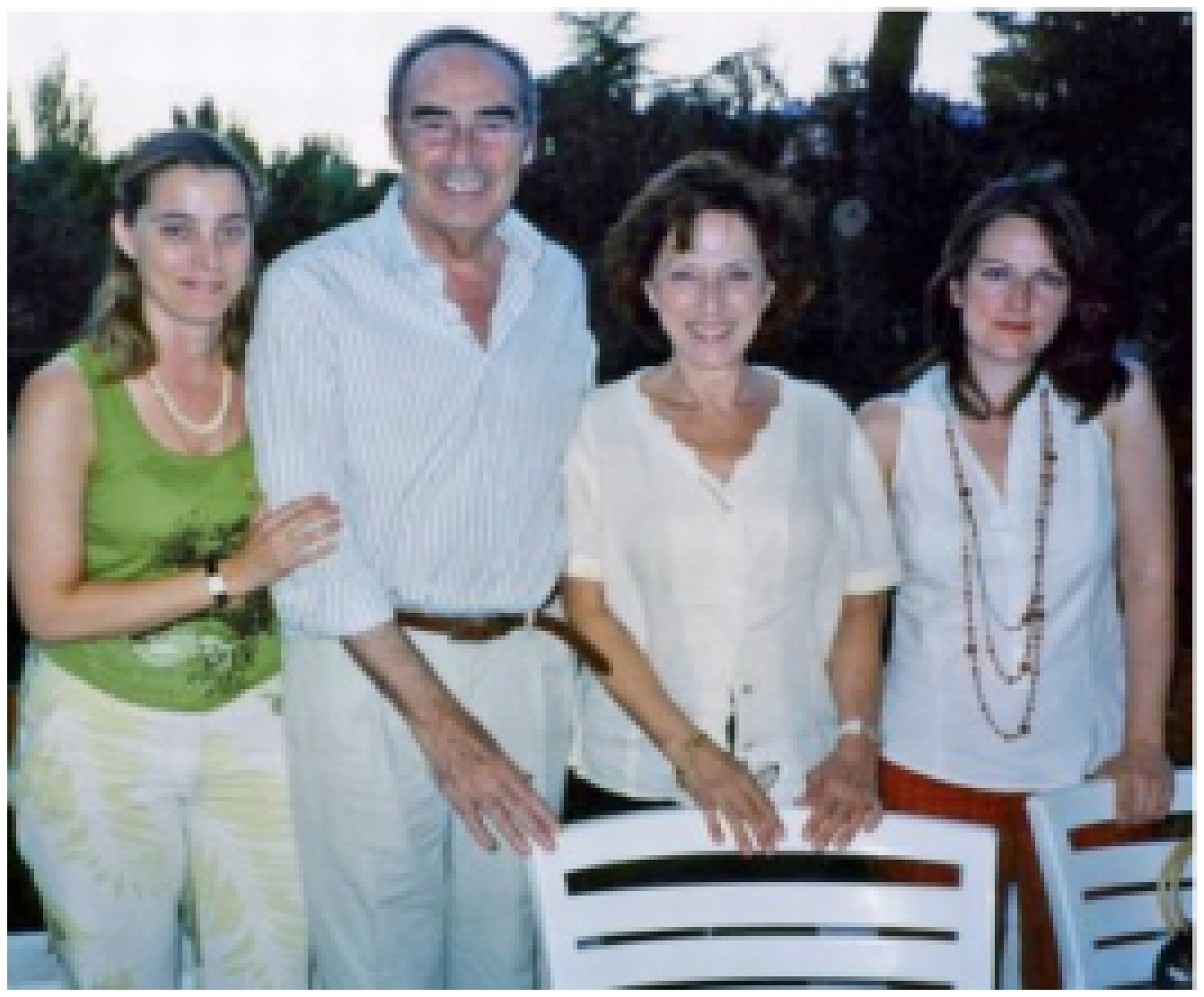}}
\end{figure}} 
\section*{Acknowledgement} 
\nin
{\small It is a pleasure to thank Elena Yndurain and Jos\'e Pelaez for communicating some of the photos, Nora Brambilla and Matthias Neubert for their invitation to present and to publish this emotional talk.}
\end{document}